\documentclass[draft]{agujournal2019}
\usepackage{url} 
\usepackage{lineno}
\usepackage[inline]{trackchanges} 
\usepackage{soul}
\linenumbers
\draftfalse

\journalname{Geophysical Research Letters}

\begin{document}
\nolinenumbers
%
%


\title{Enhancing Near Real Time AI-NWP Hurricane Forecasts: Improving Explainability and Performance Through Physics-Based Models and Land Surface Feedback}

%
%




\authors{Naveen Sudharsan\affil{1*}, Manmeet Singh\affil{1*}, Sasanka Talukdar\affil{1}, Shyama Mohanty\affil{1}, Harsh Kamath\affil{1}, Krishna K. Osuri\affil{1,2}, Hassan Dashtian\affil{1}, Michael Young\affil{1}, Zong-Liang Yang\affil{1}, Clint Dawson\affil{3,4}, L. Ruby Leung\affil{5}, Sundararaman Gopalakrishnan\affil{6}, Avichal Mehra\affil{7}, Vijay Tallapragada\affil{8}, Dev Niyogi\affil{1,3,4}}

\affiliation{1}{Jackson School of Geosciences, The University of Texas at Austin, TX, USA}
\affiliation{2}{Dept of Earth and Atmospheric Sciences, NIT Rourkela, Odisha, India}
\affiliation{3}{Cockrell School of Engineering, The University of Texas at Austin, TX, USA}
\affiliation{4}{Oden Institute for Computational Engineering and Sciences, The University of Texas at Austin, TX,USA}
\affiliation{5}{Pacific Northwest National Laboratory, Richland, WA, USA}
\affiliation{6}{Hurricane Research Division, NOAA, Miami, FL, USA}
\affiliation{7}{National Weather Service, NOAA, College Park, MD, USA}
\affiliation{8}{Environmental Modelling Center, NOAA, College Park, MD, USA}
\affiliation{*}{Equal Contribution}





\correspondingauthor{Dev Niyogi}{dev.niyogi@jsg.utexas.edu}



\begin{keypoints}
\item Artificial Intelligence-based Numerical Weather Prediction (AI-NWP) models, such as Graphcast-operational, have significantly improved hurricane track forecasting, particularly by reducing errors in mid-latitude regions.

\item Simulations with the HWRFx model demonstrate that variations in soil moisture substantially influence hurricane trajectories, underscoring the importance of land-ocean-atmosphere interactions in storm evolution.

\item Integrating land surface processes into AI-NWP models could enhance their predictive accuracy and explainability, leading to more reliable forecasts of hurricane track and intensity, particularly for landfalling storms.
\end{keypoints}

%
%

%
%


\begin{abstract}

Hurricane track forecasting remains a significant challenge due to the complex interactions between the atmosphere, land, and ocean. Although AI-based numerical weather prediction models (AI-NWP), such as Google's Graphcast operation, have significantly improved hurricane track forecasts, they currently function as atmosphere-only models, omitting critical land and ocean interactions. To investigate the impact of land feedback, we conducted independent simulations using the physics-based Hurricane WRF experimental model (HWRFx) to assess how soil moisture variations influence storm trajectories. Our results show that land surface conditions significantly alter storm paths, demonstrating the importance of land-atmosphere coupling in hurricane prediction.

Although recent advances have introduced AI-based atmosphere-ocean coupled models, a fully functional AI-driven atmosphere-land-ocean model does not yet exist. Our findings suggest that AI-NWP models could be further improved by incorporating land surface interactions, improving both forecast accuracy and explainability. Developing a fully coupled AI-based weather model would mark a critical step toward more reliable and physically consistent hurricane forecasting, with direct applications for disaster preparedness and risk mitigation.
\end{abstract}

\section*{Plain Language Summary}
Hurricanes are among the most dangerous weather events, and accurately predicting their paths is crucial to disaster preparedness. Artificial Intelligence (AI)-based weather models have made significant progress in forecasting hurricanes in recent years. However, most AI models today only consider atmospheric conditions and do not account for how hurricanes interact with the land and ocean, both of which play a critical role in storm behavior, especially near the coast.

In this study, we analyze the forecasts from Google's Graphcast-operational, an AI-based weather model, and compare them with the results from HWRFx, a physics-based hurricane model that includes land surface interactions. We show that land feedback can significantly change the path of a hurricane by running simulations with different soil moisture conditions. These findings suggest that current AI models could be improved if they incorporate land and ocean interactions into their forecasts.

While some AI models are beginning to integrate atmosphere-ocean interactions, no fully functional AI-based model couples atmosphere, land, and ocean processes. Our results highlight the potential benefits of such a system, which could provide more reliable hurricane forecasts and help communities better prepare for extreme weather events.
%
%

%


%
%
%
%

\section{Introduction}
The performance of Artificial Intelligence-based Numerical Weather Prediction (AI-NWP) models, such as Google DeepMind’s Graphcast-operational \cite{lam2023learning}, has substantially improved hurricane track forecasting, particularly in regions where traditional high-resolution NWP models, like the Weather Research and Forecasting (WRF) model, encounter difficulties \cite{rasp2020weatherbench}. These AI models have demonstrated superior skill in capturing hurricane tracks, particularly in mid-latitude regions. However, current AI-NWP models remain atmosphere-only, and do not incorporate land surface interactions, which play a crucial role in hurricane behavior, particularly during landfall \cite{kepert2018boundary}.  

Hurricane Beryl, the first storm of the 2024 Atlantic hurricane season, provides an illustrative case. Beryl rapidly intensified to a Category 5 hurricane unusually early in the season, fueled by exceptionally warm sea surface temperatures and low wind shear \cite{klotzbach2022trends}. The storm’s trajectory was complex, with multiple landfalls in Grenada, the Yucatán Peninsula, and eventually Texas, where it caused widespread flooding and power outages \cite{santini2024hurricane}. The rapid intensification and unusual track of Beryl underscore the importance of improving hurricane forecast models to capture land-ocean-atmosphere interactions accurately.

To investigate the influence of land surface interactions on hurricane forecasts, we conduct independent simulations using the multi-nested, grid-enabled Hurricane WRF experimental (HWRFx) model. Specifically, we analyze how variations in soil moisture impact storm trajectories. We use HWRFx simulations to understand how land-atmosphere feedback influences hurricane tracks. These insights highlight a key limitation in current AI-based models: the absence of land feedback, which could impact forecast accuracy.

Our results suggest that AI-NWP models could be further improved if they evolved into fully AI-driven atmosphere-land-ocean coupled systems. While recent efforts have introduced AI-based atmosphere-ocean coupled models \cite{wang2024coupled}, a fully functional AI-based atmosphere-land-ocean model has not yet been developed. This study underscores the potential benefits of integrating land surface feedback into AI-NWP models, which could lead to more physically consistent and reliable hurricane forecasts, ultimately improving disaster preparedness and response efforts \cite{bauer2015quiet}.  

This study presents a comprehensive analysis of hurricane track forecasting using Graphcast-operational, supported by soil moisture sensitivity simulations from HWRFx. By analyzing the track forecasts of hurricanes Beryl, Debby, Francine, and Helene, we explore how land surface conditions influence hurricane paths \cite{gall2013hurricane}. Our findings emphasize the need for future AI models to incorporate land-atmosphere interactions, paving the way for more advanced AI-NWP frameworks \cite{torn2012uncertainty}.

\section{Data and Methods}

\subsection{Data}
The primary dataset used to assess hurricane track accuracy in this study is the International Best Track Archive for Climate Stewardship (IBTrACS) database, which provides comprehensive hurricane track information globally \cite{kruk2010technique, jarvinen1984tropical}. This study focuses on the 2024 hurricane season, analyzing hurricanes Beryl, Debby, Francine, and Helene (Figure 1) to evaluate the accuracy of forecasted hurricane tracks.  

We also use global atmospheric data from the Global Forecast System (GFS) to provide initial conditions for the AI-based Graphcast-operational model \cite{lam2023learning} and the physics-based HWRFx model. The GFS is a global numerical weather prediction system that provides atmospheric variables, which are essential inputs for AI and physics-based weather models \cite{saha2014ncep}. The GFS system provides initial conditions at six-hour intervals, which are crucial for accurate medium-range weather forecasting \cite{hamill2013noaa}.  

We utilize soil moisture data developed by \citeA{chen2023dynamically} for simulations involving land surface interactions. This dataset was generated using an automated machine-learning-assisted ensemble framework that effectively captures the nonlinear nature of soil moisture across spatiotemporal domains, allowing us to examine how variations in soil moisture affect hurricane trajectories \cite{koster2004regions}. This data supports our investigation into critical land-atmosphere feedback mechanisms in hurricane prediction \cite{santanello2018land}.  

\begin{figure}
\includegraphics[scale=0.6]{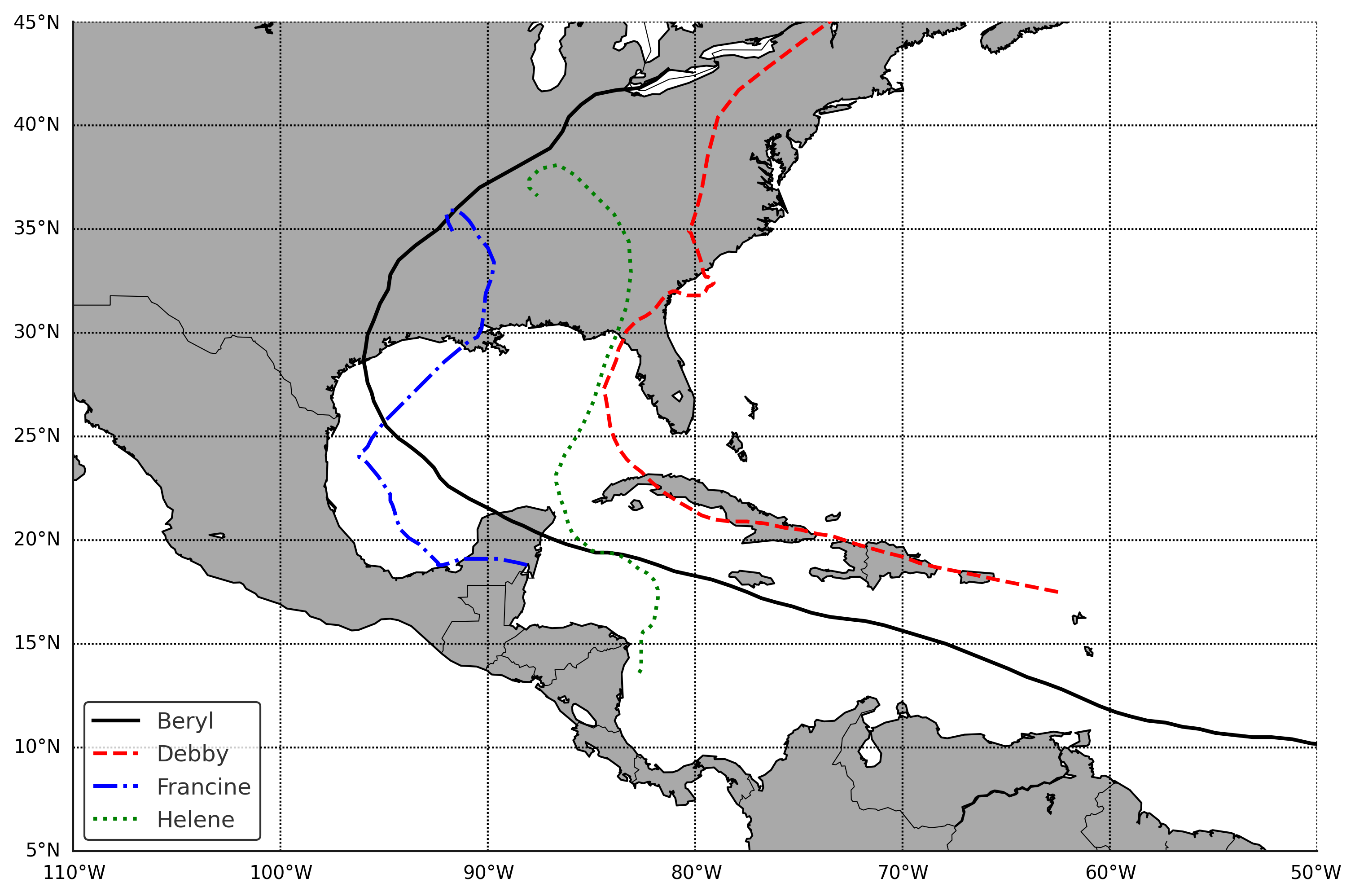}
\caption{Best tracks for hurricanes Beryl, Debby, Francine, and Helene from IBTrACS.}
\label{fig:Figure 1}
\end{figure}

\subsection{Models}

\subsubsection{Graphcast-operational model}
Graphcast-operational is a state-of-the-art AI-based Numerical Weather Prediction (NWP) model that employs deep learning to produce medium-range weather forecasts \cite{lam2023learning}. The model, trained on four decades of historical weather data from the ECMWF ERA5 dataset, can generate 10-day forecasts with high accuracy \cite{hersbach2020era5}.  

For this study, we use Graphcast-operational to simulate the tracks of hurricanes Beryl, Debby, Francine, and Helene. The model operates by processing spatially structured data using Graph Neural Networks (GNNs) and predicts multiple Earth-surface and atmospheric variables at multiple levels across a global grid with a resolution of 0.25 degrees latitude/longitude \cite{battaglia2018relational}. Compared to traditional NWP models like WRF, Graphcast-operational significantly reduces computational time and provides more accurate hurricane track predictions, particularly in mid-latitude regions where conventional models exhibit larger forecast errors \cite{weyn2020improving}.  

The output from Graphcast-operational includes forecasts of hurricane tracks and intensities in terms of 10-meter maximum wind speed and mean sea level pressure (MSLP). These outputs were then compared to the observed hurricane tracks from the IBTrACS database \cite{kruk2010technique}. Furthermore, we performed multiple simulations using different initialization times based on the GFS initial conditions to evaluate how forecast accuracy evolved over time \cite{saha2014ncep}.  

\subsubsection{HWRFx Model}
To explore the role of land surface processes in hurricane forecasting, we conduct simulations using the Hurricane Weather Research and Forecasting Experimental (HWRFx) model. HWRFx, deployed at the University of Texas at Austin, extends severe weather forecasting capabilities beyond standard HWRF implementations.  

The traditional HWRF model, developed by NOAA, is designed for hurricane simulations using two telescopic, high-resolution, two-way movable nested grids embedded in a static parent domain optimized for hurricane physics \cite{gopalakrishnan2011experimental}. Further developments led to the creation of basin-scale HWRF, which incorporates up to five moving nests to track multiple tropical cyclones (TCs) in a single simulation, enabling high-resolution multi-TC forecasts \cite{zhang2016increasing, alaka2022high}.  

HWRFx builds upon these advancements by integrating the WRF Nonhydrostatic Mesoscale Model (NMM) dynamical core, advanced flexible physics from WRF, and high-resolution nesting techniques from Basin-scale HWRF. It introduces a dual-nesting approach, employing both static and moving nests in a single simulation. Static grids focus on regions with severe weather, while dynamic moving nests track storms such as hurricanes and monsoon depressions, allowing for real-time high-resolution monitoring and adaptive simulation.  

HWRFx also supports ocean-atmosphere coupling via the Hybrid Coordinate Ocean Model (HYCOM), enabling seamless interaction between the atmosphere and ocean to enhance forecast accuracy. The system deploys high-resolution nests (1 km and 3 km) within a parent domain of 9 km resolution, optimizing computational efficiency while resolving synoptic to convective scales.  

This modeling system has been rigorously tested for hurricanes, monsoon depressions, heavy rainfall, and severe thunderstorms. It demonstrates stable performance even with up to 11 high-resolution nests within a continental-scale parent domain. Compared to operational HWRF, HWRFx offers greater flexibility by integrating dual-nesting techniques with advanced ocean-land-atmosphere coupling. While operational HWRF primarily focuses on real-time hurricane forecasting, HWRFx provides a more versatile framework for severe weather simulations.  

This study used HWRFx to assess the impact of soil moisture variations on hurricane tracks. We modified initial soil moisture conditions based on data from \citeA{chen2023dynamically} and conducted multiple simulations to evaluate how different land surface conditions influence storm paths.

\subsection{Analysis}
\subsubsection{Hurricane Track Evaluation}
To assess the accuracy of hurricane track predictions from both Graphcast-operational and HWRFx, we calculated the deviation of the predicted storm paths from the observed tracks recorded in the IBTrACS database. Forecast errors were analyzed across multiple initialization times at six-hour intervals, comparing predicted positions with actual storm locations for up to five days. The forecast error at 30 hours lead time was used as a reference for evaluating five-day track errors. The great-circle distance metric was used to quantify deviations between forecasted and observed storm centers.  

\subsubsection{Soil Moisture Sensitivity}
To quantify the impact of soil moisture on hurricane trajectories, we conducted a series of HWRFx simulations with varying initial soil moisture conditions. The simulations explored both increased and decreased soil moisture relative to baseline conditions, spanning from wilting point (WP) to field capacity (FC). The storm tracks were then calculated for each scenario.   

\section{Results}
\subsection{Graphcast-operational Hurricane Track Forecasting Results}

The performance of the Graphcast-operational AI model was evaluated by simulating the tracks of hurricanes Beryl, Debby, Francine, and Helene. Compared to the physics-based model HWRFx, Graphcast-operational demonstrated significant improvements in predicting hurricane trajectories, particularly in mid-latitude regions, where traditional models tend to exhibit larger forecast errors.  

\begin{figure}
\includegraphics[scale=0.45]{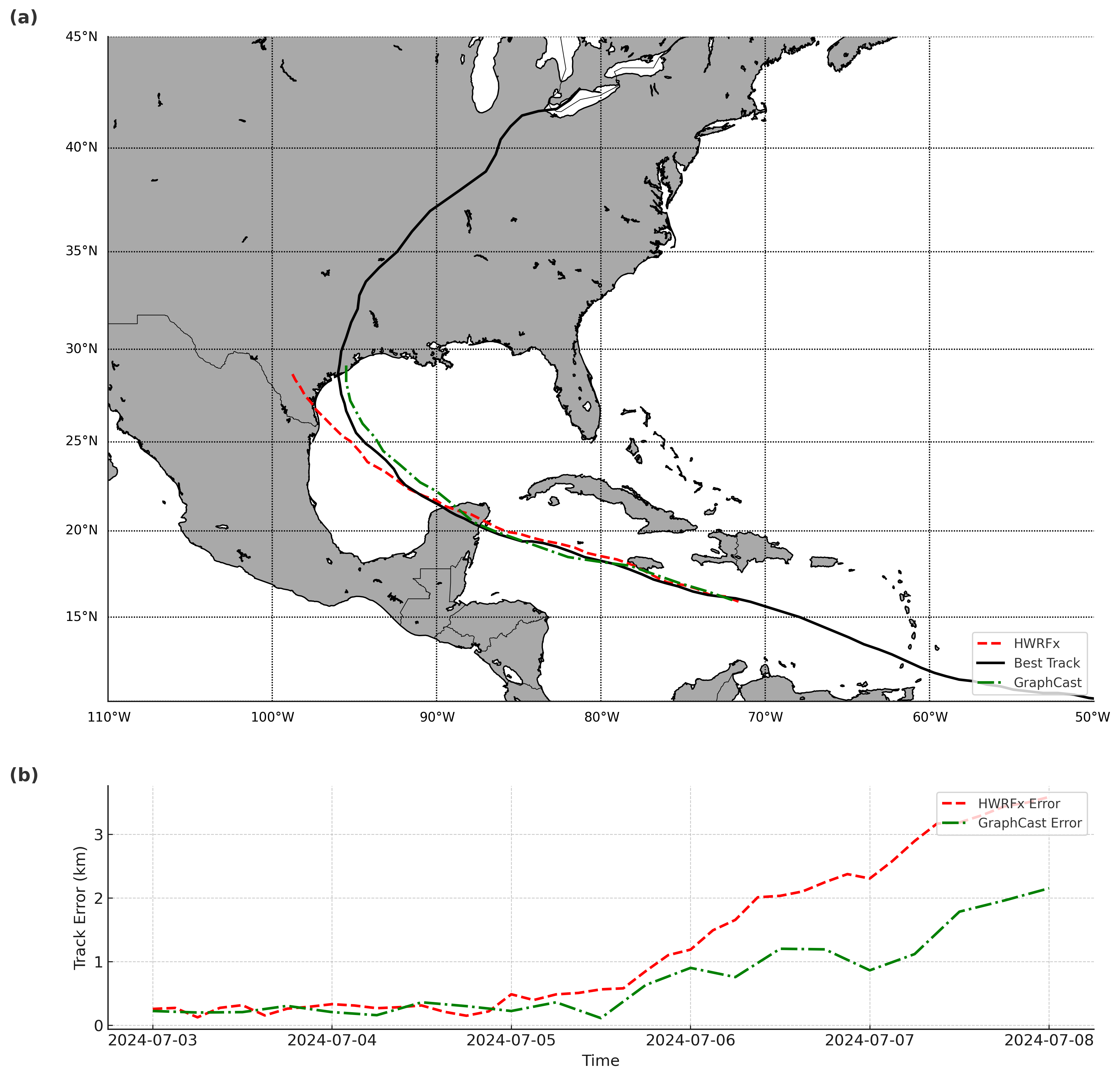}
\caption{The forecasted tracks for Hurricane Beryl from HWRFx and Graphcast-operational with an initial condition of 07-03-2024 00hrs. Figure (a) compares tracks from HWRFx and Graphcast-operational with the best tracks from IBTrACS. The forecast is for 5 days, valid until 07-08-2024 00hrs. Graphcast-operational closely reproduces the best track, but the landfall timing differs. Figure (b) shows track errors for both models. While both models perform well for the first 3 days, Graphcast-operational outperforms HWRFx beyond this timeframe.}
\label{fig:Figure 2}
\end{figure}

For Hurricane Beryl, Graphcast-operational accurately predicted the storm's trajectory as it moved from the Atlantic toward the Caribbean. The track error over a 5-day forecast period was reduced by approximately 40\% compared to HWRFx (Figure \ref{fig:Figure 2}). Similarly, for Hurricanes Debby (Figure S1), Francine (Figure S2), and Helene (Figure S3), Graphcast-operational showed a reduction in track error of 35\%, 30\%, and 37\%, respectively, on the 5th day, demonstrating consistent performance across multiple storm systems. These results highlight the superior predictive skill of Graphcast-operational, especially in capturing long-term storm trajectories.  

In addition to improved track accuracy, Graphcast-operational also produced more reliable forecasts for central pressure throughout the hurricanes’ life cycles. The AI-based model successfully captured the timing of rapid intensification phases, with the forecast closely aligning with observed data from the National Hurricane Center (Figure S4). These findings underscore the potential of AI-based models in operational hurricane forecasting, providing critical lead times for preparedness and disaster management.  

\subsection{HWRFx Simulation Results: Impact of Land Surface Processes}

\begin{figure}
\includegraphics[scale=0.6]{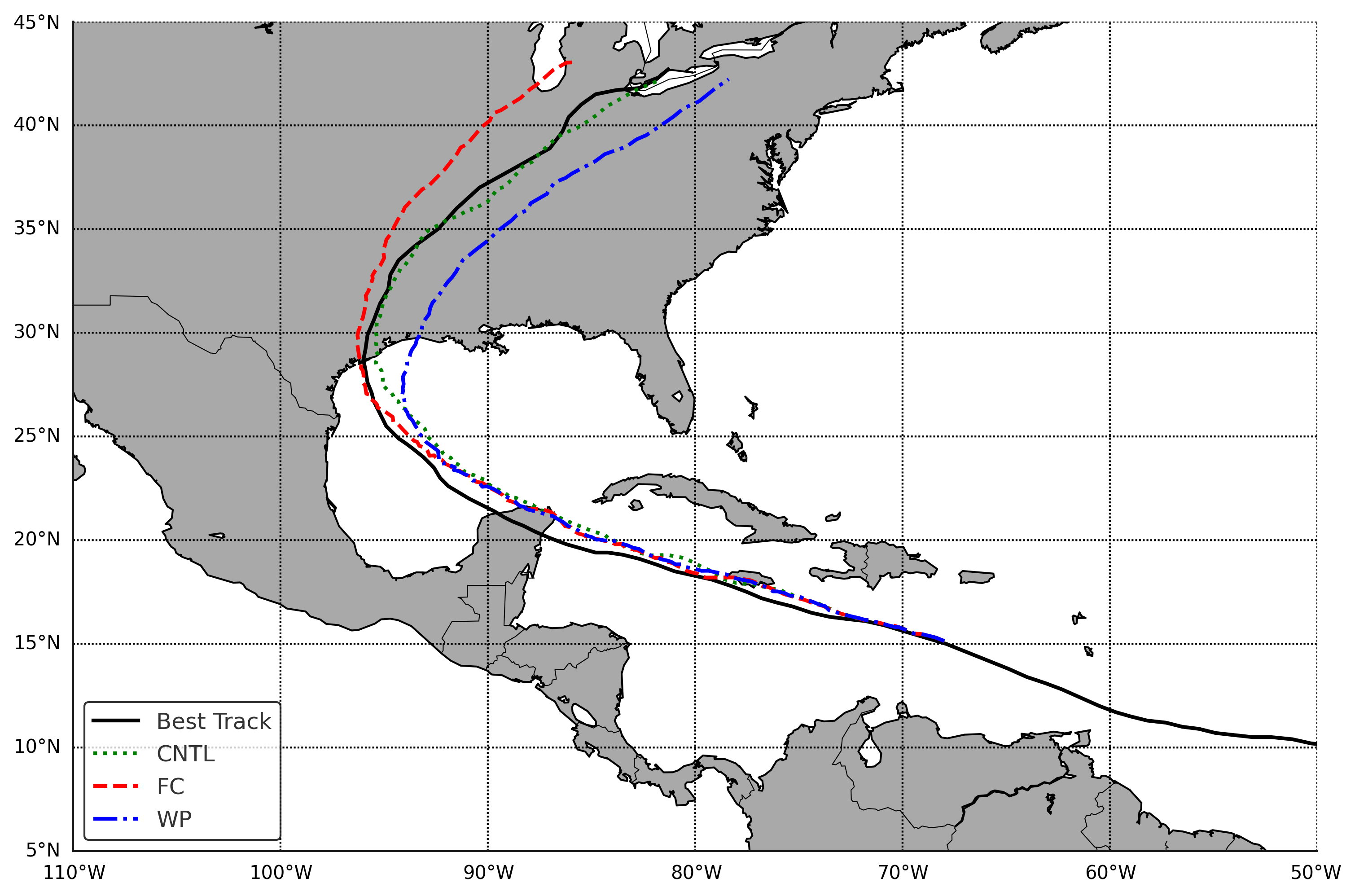}
\caption{Simulated tracks of Hurricane Beryl under different soil moisture conditions using HWRFx. The "Best Track" (dashed blue line) represents the observed hurricane path. The control simulation (CNTL) with default soil moisture is shown in black, while the Field Capacity (FC) and Wilting Point (WP) scenarios are represented in blue and red, respectively. The results indicate a westward shift in the FC storm track and an eastward shift in the WP storm track. Notably, all tracks show strong agreement over the ocean but diverge significantly as they approach land (~350 km from the Gulf of Mexico coastline), demonstrating the influence of land-atmosphere interactions.}
\label{fig:Figure 3}
\end{figure}

Figure \ref{fig:Figure 3} presents the results of Hurricane Beryl's track simulations under different soil moisture conditions, as modeled by HWRFx. The dashed blue line represents the "Best Track", or the observed hurricane path from the National Hurricane Center. The control simulation (CNTL), shown in black, maintains default soil moisture conditions. The Field Capacity (FC) scenario (blue line) represents increased soil moisture, while the Wilting Point (WP) scenario (red line) represents decreased soil moisture.  

In the control run, HWRFx initially predicted a trajectory similar to the observed track but began to deviate near the Gulf Coast, ultimately forecasting landfall farther west than the best track, then after applying vortex correction, the control run has shown similar track as best track but east of the track most of the time.  In the Field Capacity (FC) simulation, increased soil moisture shifted the hurricane’s path significantly westward. This suggests that higher soil moisture alters the storm’s trajectory, likely due to enhanced frictional and heat exchange processes. Conversely, the Wilting Point (WP) simulation, which represents drier soil conditions, shifted the storm’s path eastward, suggesting that drier land reduces surface friction, allowing the storm to maintain a more eastward trajectory and higher intensity for an extended period.  

Despite these insights, HWRFx did not accurately reproduce Hurricane Beryl’s landfall in any scenario. Additionally, near real-time HWRFx simulations initialized 5 days in advance predicted a westward track, while Graphcast-operational accurately forecasted landfall near Houston. This discrepancy highlights the challenges of traditional physics-based models in representing storm-land interactions, reinforcing the importance of improving AI-based models with land-ocean-atmosphere coupling.  

\subsection{Future Directions: Fully AI-Coupled Atmosphere-Land-Ocean Models}

\begin{figure}
\includegraphics[scale=0.12]{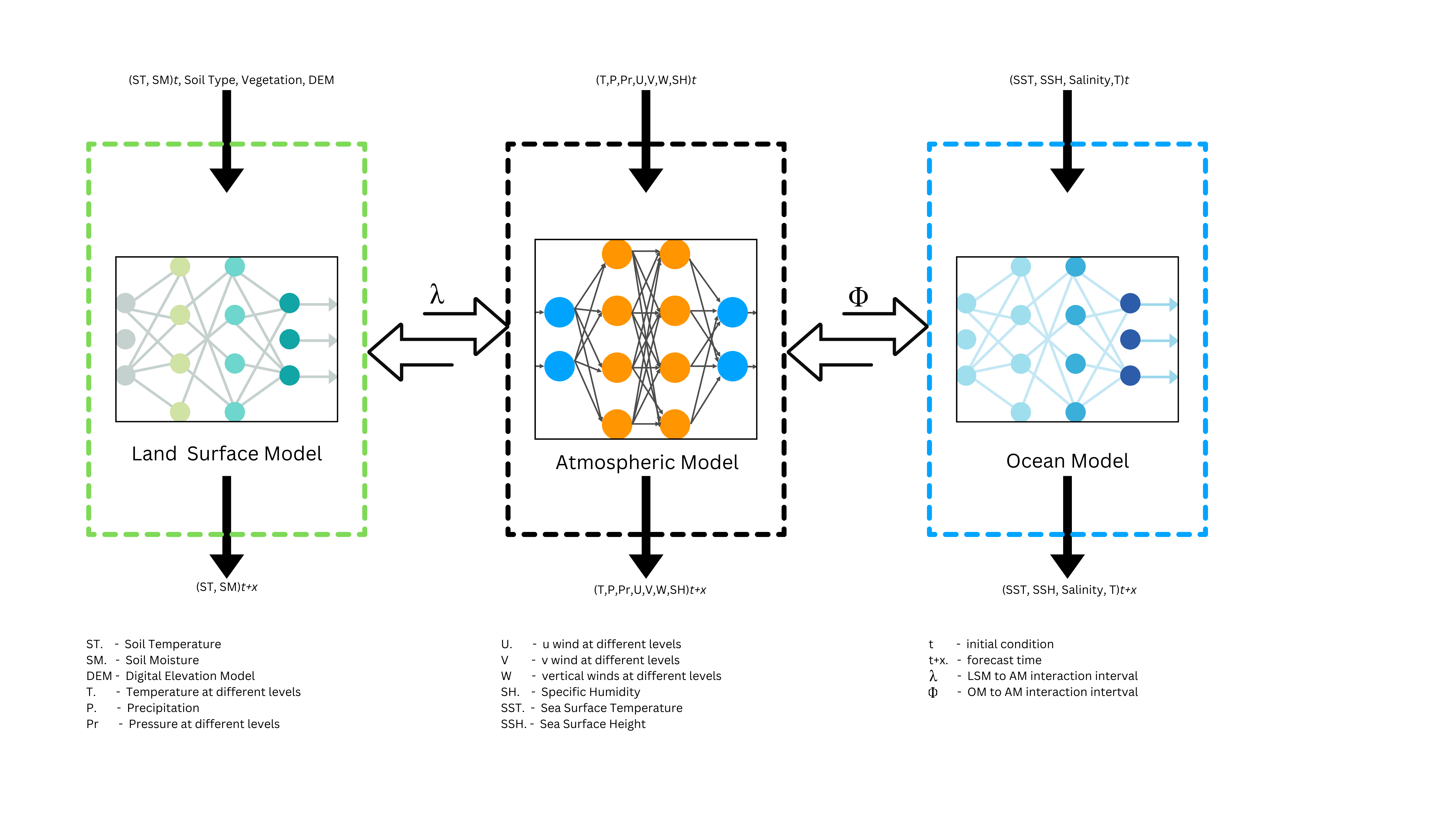}
\caption{Schematic of a proposed AI-based coupled model incorporating land and ocean processes into an atmospheric AI forecasting system.}
\label{fig:Figure 4}
\end{figure}

Figure \ref{fig:Figure 4} illustrates a conceptual framework for integrating land, ocean, and atmospheric processes into an AI-NWP model. Current AI-NWP models, including Graphcast-operational, primarily focus on atmospheric processes and, in later studies, atmosphere-ocean interactions, but they do not fully account for land surface feedback mechanisms, which play a pivotal role in hurricane evolution.  

As hurricanes approach land, variations in soil moisture, surface heat flux, and terrain-induced frictional effects can significantly alter their intensity and trajectory. The results of this study suggest that incorporating land surface feedback into AI-NWP models could significantly improve their forecasting capabilities and explainability. This integration would allow AI-based models to provide more reliable hurricane forecasts, particularly for landfalling storms, where land-atmosphere feedbacks become dominant in storm evolution.  

Future AI-NWP models should evolve beyond purely atmosphere-based frameworks into fully AI-driven coupled atmosphere-land-ocean systems. This approach would enhance forecast accuracy, lead times, and physical interpretability, ultimately improving operational hurricane prediction and disaster preparedness strategies for coastal regions.  

\section{Discussion and Conclusions}

This study demonstrates the significant potential of AI-based models, particularly Graphcast-operational, in improving hurricane track forecasts, especially in regions where traditional models like WRF tend to struggle. By analyzing the hurricane tracks of Beryl, Debby, Francine, and Helene we found that Graphcast-operational consistently produced more accurate track forecasts with lower errors and longer lead times. These improvements are particularly notable during mid-latitude transitions and rapid intensification phases, where traditional physics-based models often exhibit larger forecast errors. Enhancing track predictions in these scenarios is critical for operational forecasting, as it provides decision-makers with more time to prepare and respond to impending hurricanes.  

Beyond the advancements in AI-based forecasting, this study highlights the critical role of land surface interactions, particularly soil moisture, in hurricane dynamics. Through a series of independent HWRFx simulations, we demonstrated that variations in soil moisture significantly alter storm trajectories. Our findings indicate that increased soil moisture enhances surface friction and cooling effects, leading to a trajectory shift towards West. Conversely, reduced soil moisture results in drier, less frictional land surfaces, allowing storms to shift eastward. These results underscore the importance of land-atmosphere interactions, which remain absent from current AI-based models but could play a crucial role in improving hurricane forecasts.  

Despite its ability to simulate land interactions, the HWRFx model exhibited notable track deviations from the best track when run near-real time.  Physics-based models like HWRFx forecasted a trajectory which was westward of the best track five days in advance, whereas Graphcast-operational correctly predicted landfall near Houston. This discrepancy highlights the limitations of traditional physics-based models in capturing the complexities of land surface feedback under rapidly evolving atmospheric conditions. At the same time, the ability of Graphcast-operational to maintain high forecast accuracy, even at long lead times, emphasizes the growing potential of AI-based approaches in operational hurricane forecasting.  

The schematic representation of ocean-atmosphere-land feedback further emphasizes the need for fully coupled AI-NWP models that integrate land surface processes alongside ocean and atmospheric dynamics. While some recent AI models have incorporated atmosphere-ocean interactions, a fully functional AI-based atmosphere-land-ocean coupled model has not yet been developed. Our findings suggest that incorporating land surface feedback into AI-based forecasting systems could significantly enhance both prediction accuracy and explainability, particularly for landfalling storms where land interactions become a dominant factor in storm evolution.  

Overall, this study highlights the importance of integrating land surface processes into AI-based weather models to improve hurricane track forecasts. As hurricanes continue to pose increasing risks due to climate change, advancing AI-NWP models beyond their current atmosphere-only framework is critical. The inclusion of land-ocean-atmosphere interactions in AI models could lead to more reliable hurricane forecasts, ultimately improving disaster preparedness, mitigation strategies, and resilience for vulnerable coastal communities. These findings suggest that future advancements in AI-based forecasting should prioritize fully AI-driven coupled atmosphere-land-ocean models, bridging the gap between data-driven approaches and physically consistent numerical weather predictions.

\section*{Open Research Section}

\section*{Data Availability Statement}
The data used in this study are available from the following sources: Hurricane track data for the 2024 season was obtained from IBTrACS database (\url{https://www.ncei.noaa.gov/products/international-best-track-archive}). The initial conditions for the AI-based Graphcast-operational model were provided by the Global Forecast System (GFS), accessible through the National Centers for Environmental Prediction (NCEP) (\url{https://www.ncdc.noaa.gov/data-access/model-data/model-datasets/global-forcast-system-gfs}). Soil moisture, temperature, and atmospheric parameters for the HWRFx simulations were sourced from the European Centre for Medium-Range Weather Forecasts (ECMWF) ERA5 reanalysis dataset (\url{https://www.ecmwf.int/en/forecasts/datasets/reanalysis-datasets/era5}). All the necessary simulation outputs from Graphcast-operational and HWRFx used in this study are available upon request from the corresponding author.

\acknowledgments
The authors thank the original developers of Graphcast for releasing the model weights and code that helped us set it up and run the inference. The authors also gratefully acknowledge the use of the ADCIRC and ATM23014 allocation at the Texas Advanced Computing Center at the University of Texas at Austin. This work was supported by the United States Department of Energy under grant DE-SC0022211 - MuSiKAL: Multiphysics Simulations and Knowledge discovery through AI/ML technologies.

%
%


%
%
%
%
%

\bibliography{agujournaltemplate}




\end{document}